\documentclass[12pt]{article}
\usepackage{graphics}
\usepackage[dvips]{graphicx}
\usepackage{listings}
\begin{document}
{\footnotesize jcis@epacis.org}
\begin{center}

{\bf Use of Python programming language in astronomy and science}
\bigskip

{\small Daniel M. Faes\footnote{E-mail Corresponding Author: moser@usp.br}\\ 
}
\smallskip
{\small
{Instituto de Astronomia, Geof\'isica e Ci\^encias Atmosf\'ericas, Universidade de S\~ao Paulo, Rua do Mat\~ao 1226, Cidade Universit\'aria, 05508-900, S\~ao Paulo, SP, Brazil}\\
}
{\footnotesize Received on September **, 2011 / accepted on *****, 2011}
\end{center}

\begin{abstract}
The use of Python is noticeably growing among the scientific community, and Astronomy is not an exception. The power of Python consists of being an extremely versatile high-level language, easy to program that combines both traditional programming and data reduction and analysis tools. Here I make a brief introduction to Python, mentioning a few programming practices implemented in the language and some of its useful features on the process of data manipulation. I cover in a little more detail the standard scientific libraries (NumPy and SciPy) for data handling, the graphical library (Matplotlib), and tools for specific use in astronomy (PyFITS and PyRAF). Good programming practices and how they are applied at the language are also viewed. Python resources and references are mentioned throughout the text for those who wish to go deeper and make use of the power of the language. 
\bigskip

{\footnotesize
{\bf Keywords}: Programming language: Python, Data analysis and visualization, Computing resources.}
\end{abstract}

\textbf{1. INTRODUCTION}

\bigskip
\bigskip
In short, Python is a free, cross-platform, widely used and well documented programming language. Since it is a multi-purpose language, designed to have extensive interaction with other languages, it can integrate different stages of a process of data manipulation, since the reduction, visualization and integration with databases. This is one of the reasons of the growing use of this language. 

Python already has a significative use inside the scientific community. I mention some recent publications show that. We see Python at computing applications and software (\cite{2007PLW} and \cite{2007ASPC..376..261M}), at pure physics ones (\cite{2011JHEP...06..128A} and \cite{2007JOSAA..24.1580H}) and astronomical publications (\cite{2010A&A...524A..61N} and \cite{2008PASP..120..439C}). Among them, we can see Python contributing in different process of the scientific production.

A typical example of a project that makes uses of Python at the $Sage$ project \cite{sage}. $Sage$ is a free open-source mathematics software system that combines many existing open-source packages into a common Python-based interface.

Other concrete example of the mentioned Python features is the project $PyNBody$, an collaborative toolkit for astrophysical N-body simulations, focused on star clusters and galactic centers \cite{pynbody}.
Since the project philosophy, based on collaborative development, it uses Python to implement the scientific problem, integrating codes of different languages (e.g., $C/C++$), making use of GPU's ($OpenCL$ and $OpenGL$ codes), to parallel calculation and visualization resources.

Python is also present on common used data manipulation platforms. A general scientific one is the Enthought Python Distribution \cite{enthought}. 
At astronomy, Python is present at the European Southern Observatory Astronomical Software Collection, $scisoft$ \cite{scisoft}, at Space Telescope Science Institute software \cite{STScI}, among others.

In this article a quick start to Python is made at section 2, where a little about its syntax is discussed; at section 3, the standard scientific and graphical tool are showed. The main astronomical libraries are present at section 4. Some important programming practices on Python are at section 5, and our conclusions are present at section 6.

\bigskip
\bigskip

\textbf{2. STARTING WITH PYTHON}

\bigskip
\bigskip
Similar to other computational languages, Python can be programmed in static or interactive way. Thus, the memory management is made automatically, and it may be inefficient for large amounts of data. About Python integration with other low-level languages, see section 6.

Interactive programming is very useful when developing a new code or trying to debug one. There are three built-in ways of programming at Python: i) the basic interpreter $python$, ii) the enhanced interpreter $ipython$ and iii) the integrated development environment $IDLE$. There is also other proprietary software to user choice.

Beyond the standard $python$ console, the $ipython$ offers a friendly programming environment, with integrated system shell commands (e.g., $ls$, $cd$, etc). Also you can have full system access using the prefixes $!$ and $!!$ to commands. It lists the options associated with the command or variable with through the keyboard key $Tab$. As last point to mention, it numerates the commands and color specific sections of the Python messages.

$idle$ is the development environment which integrates the editor to the Python console. Many features of the $ipython$ can also be found at environment.

The static programming can be made at any text editor, keeping the definition of filename extension of $*.py$. Then, the program can be executed calling the Python executable interpreter with the program filename in sequence. This illustrates the versatility of language: there is no need for compilations - the same code runs directly on various platforms. It just is not true if you choose to use functions applied to specific operating systems.

The program (called sometimes as script) can be self-executing. An example at Unix of how to make it is simply add the following line at the beginning of the file:
\begin{lstlisting}
#!/usr/bin/env python
\end{lstlisting}
We are invoking the command $env$, which seeks the path of the command passed as an argument (i.e. $python$) in the environment variables in order to run it. Remember that the file must have execution permission.

About Python syntax, it is very similar to other languages. But a key structure at Python is the attributes, or ``proper functions" set to the Python variable type or rather, $object$ or $class$. Its syntax is defined as a point after the variable.

\begin{lstlisting}
>>> #<object_name> + point + <object_attribute>
>>> wrong = 'ex@mple'
>>> right = wrong.replace('@','a')
>>> print(right)
example
>>> #In this case, 'replace' is an attribute of strings.
\end{lstlisting}

Besides the attributes embedded in language, external library functions (``modules") can be invoked. They are called exactly the same way that attribute does.
\begin{lstlisting}
>>> #import numpy library with alias np
>>> import numpy as np
>>> print(np.pi)
3.14159265359
>>> avg = np.average([0.1,1,10])
>>> print(avg)
3.7
\end{lstlisting}

We can see more about the properties of variables and modules with the command help.
\begin{lstlisting}
>>> #About the built-in variables:
>>> help(float)
>>> #About an imported function (as np.pi):
>>> help(np.pi)
\end{lstlisting}
\bigskip
\bigskip

\textbf{3. THE STANDARD LIBRARIES}

\bigskip
\bigskip
The mostly disseminated Python libraries are the numerical tools NumPy with SciPy \cite{scipy} and graphical library Matplotlib \cite{matplotlib}. At this section there is a brief description of them.

NumPy is the Python numerical tool designed to work with large vectors and multi-dimensional arrays, a practical manner and maintaining high performance.
Between NumPy functions are the basic mathematical functions ($log$, $exp$, $min$, $max$...), trigonometric ($sin$, $cos$, ...), Bit-twiddling, matrix calculation, the discrete Fourier transform ($numpy.fft$) , linear algebra ($numpy.linalg$), random samples ($numpy.random$) and others.

\begin{lstlisting}
>>> import numpy
>>> x = numpy.arange(0, 2*np.pi, 0.1)
>>> print(len(x))
63
>>> y = np.sin(x)
>>> print(numpy.max(y))
0.999573603042
\end{lstlisting}

A remarkable feature of NumPy is its file I/O capabilities, through the $numpy.loadtxt$ and $numpy.savetxt$ functions. 
\begin{lstlisting}
>>> import numpy
>>> #Creating a 4x3 matrix in numerical sequence
>>> m = numpy.arange(12).reshape(4,3)
>>> print(m)
[[ 0  1  2]
 [ 3  4  5]
 [ 6  7  8]
 [ 9 10 11]]
>>> numpy.savetxt('matrix.dat', m, delimiter='\t',\
... fmt='%2.1f')
>>> m = []
>>> m = numpy.loadtxt('matrix.dat')
print(m)
array([[  0.,   1.,   2.],
       [  3.,   4.,   5.],
       [  6.,   7.,   8.],
       [  9.,  10.,  11.]])
\end{lstlisting}

The SciPy library provides many user-friendly and efficient numerical routines of scientific applications, such as routines for numerical integration ($scipy.integrate$) and optimization ($scipy.interpolate$, $scipy.optimize$), statistcs ($scipy.stats$), multi-dimensional image processing ($scipy.ndimage$), signal processing ($scipy.signal$), among others.

\begin{lstlisting}
>>> import numpy
>>> import scipy.integrate
>>> #Creating the function f(var)=var^2
... def f(x): 
...   return x**2
...
>>> #Defining an interval and its resolution
... x = numpy.linspace(1,3,10)
>>> #Calculating the integral by different methods
... scipy.integrate.quad(f, 1, 3)
(8.666666666666666, 9.621932880084691e-14)
>>> scipy.integrate.trapz(f(x), x)
8.6831275720164616
>>> scipy.integrate.simps(f(x), x)
8.6684956561499771
>>> #Which method is 'quad'?
... help(scipy.integrate.quad)
\end{lstlisting}

Matplotlib is a Python plotting library which produces publication quality figures in a variety of formats and even interactive environments across platforms. As a Python library, Matplotlib can be used in scripts, the $python$ and $ipython$ shell and even web application servers.

\begin{lstlisting}
>>> #Remember: this definitions are equivalent!
... import pylab
>>> #Or
... import matplotlib.pyplot
>>> import numpy
>>> #A graphical SciPy example of interpolation
... import scipy.interpolate
>>> x = numpy.linspace(0, 10, 7)
>>> y = numpy.exp(1./3*x)
>>> f_intp = scipy.interpolate.interp1d(x, y)
>>> x_intp = numpy.linspace(0, 10, 100)
>>> #Plotting the original points (blue) and \
... # the interpolated line (green)
... pylab.plot(x, y, 'bo')
>>> pylab.plot(x_intp, f_intp(x_intp), 'g')
>>> pylab.show()
>>> #A more sophisticated interpolation...
>>> f_spl = scipy.interpolate.UnivariateSpline(x,y)
>>> pylab.plot(x_intp, f_spl(x_intp), 'r')
>>> #An extrapolation outside points limits
>>> x_extra = numpy.linspace(10, 11, 10)
>>> pylab.plot(x_extra,f_spl(x_extra),'m:')
\end{lstlisting}

Matplotlib is an excellent utility for graphical programs. It can be used with in $Gtk+$ applications, as well in $Qt4$ ones, including $WxWidgets$. An good reference for graphical programs with Matplotlib is \cite{tosi}.

\bigskip
\bigskip

\textbf{4. THE ASTRONOMICAL LIBRARIES}

\bigskip
\bigskip

Python has support for dealing with FITS image manipulation, and an integration with the $IRAF$ platform, resources largely used at astronomy \cite{STScI}. So, once you use Python on this process, all the old scientific resources are still available, with all the power of the language. In this case, I would say just one feature that justifies its use: its ability of dealing with errors and exceptions!

PyFITS is the module that lets you manipulate $fits$ images in Python. By it we have access to reading and editing of all information available in the format, with the possibility of being treated with Python resources.

\begin{lstlisting}
>>> import pyfits
>>> #Opening a fits image
... imname = 'ugc1437.fits'
>>> imfits = pyfits.open(imname)
>>> #Showing image header
... print(imfits[0].header)
>>> #Defining the header as an object
... imhdr = imfits[0].header
>>> #Adding information to it.
... imhdr.add_history('IWCCA PyFits example')
>>> print imhdr.ascardlist()[-1]
>>> #Looking for an especific tag
... print(imhdr.ascardlist().keys())
>>> #And making a change
... imhdr.update('OBSERVER', 'Edwin Hubble')
>>> #Acessing directly the data values
... imdata = imfits[0].data
>>> #Looking at the values of a particular section
... imdata[30:35, 10:15]
>>> #Now some juggling with the data
... import numpy as np
>>> #Let's see the minimum on all data
... np.min(imdata)
>>> #And subtract it
... imdata = imdata-np.min(imdata)
>>> #We can save an image with these changes
>>> pyfits.writeto ('out.fits' imdata, imhdr)
\end{lstlisting}

PyRAF is handling all the power of Python integrated with IRAF. Can be used by as and environment (command $pyraf$ at system shell) or as a Python module. At the PyRAF environment almost all the functionality of the $cl$ is available. Even $cl$ scripts run in PyRAF (of course, it will be much easier to debug this scripts!). The entire environment is at a more powerful graphics system, including the new GUI parameter editor.

\begin{lstlisting}
>>> from pyraf import *
>>> #Running the IRAF imstat \
... # for a Python list of names
... a = ['image1.fits','image2.fits','image3.fits']
>>> for i in range(len(a)):
...    iraf.imstat(a[i])
...
\end{lstlisting}

It is possible to mention many astronomical tools that are being develop in Python. A typical example that makes use of the graphical features of Python is the Astronomical Plotting Library in Python (APLpy) \cite{aplpy}, a module that aims to produce high quality astronomical charts. It is tutorial shows very well its capabilities.

In the area of broad growth of VOs, Python is being widely used. Just to mention one good example, it is AstroGrid Python \cite{astrogrid}, designed to write Python scripts to call VO services and construct workflows.

\bigskip
\bigskip

\textbf{5. PROGRAMMIMG PRACTICES}

\bigskip
\bigskip
Looking at codes traditionally employed on scientific programming, it is very easy to see very ugly and confusing codes! It is really a big program when we are facing a complex code, even more if the code is intend to not be develop by a single person (as the case of all collaborative projects). So Python can make very useful contributions to the readability of the codes and consequently to its maintenance problems.

At this section we mention some of good programming practices available to Python users, tools to make the codes functional in all their aspects, as creation, readability and efficiency.

The first and very important definition is the code style. In contrast to many traditional codes, Python has no field delimiters: they are controlled by indentation. There is no standard format for spacing, but the user must be consistent! The user is strongly encouraged not to use the $Tab$ key, but instead use 4 spaces. This, and other style definitions for the code main text are described at Python Enhancement Proposals number 8 (PEP 8) \cite{PEP}.

Python has many ways of structuring data, some extremely important for a good code. We strongly encourage the user to know the Lists as a Queues, the List Comprehensions, Dictionaries, and the (nested) Tuples and Sequences, or the possibility of variables from more than one type of data, with variables within variables. Working with binary saved data (e.g., $pickle$ library) can be an excellent way of optimization.

Something that is always underestimated by developers is documentation. To make it a very useful feature, see PEP 257 for Docstrings \cite{PEP}. Of course this will be as important as the code size, as useful as structure data and functions as a Python Class, or invoke functions as modules.

\begin{lstlisting}
>>> #Creating a class: BaseClass
... #""" are the docstring signal... see PEP 257!
... class BaseClass:
...   """ This data will exist in all
...   BaseClasses (even uninstantiated ones)"""
...   Name = 'BaseClass'
...   def __init__(self, arg1, arg2):
...       """ __init__ is a class constructor
...       __****__ is usually a special class method.
...       These values are created
...       when the class is instantiated."""
...       self.value1 = arg1
...       self.value2 = arg2
...   def display(self):
...       """ Self is used as an argument to
...       pretty much all class functions.
...       However, you do NOT need to pass
...       the argument self if you call this method
...       from a Class, because the class provides
...       the value of itself."""
...       print(self.Name)
...       print(self.value1+' '+self.value2)
... 
>>> obj = BaseClass('Hello','World!')
>>> obj.display()
BaseClass
Hello World!
>>> #Change one of the class parameters
... obj.value2 = 'I WCCA!!!'
>>> obj.display()
BaseClass
Hello I WCCA!!!
>>> #See the docstring!
... help(obj)
\end{lstlisting}

Finally, Python operating system interfaces are marvelous resources (probably many will not want to know about shell scripts anymore – see $os$ and $shutil$ libraries). 

For whom are concerned in use Python as a language of high computational efficiency, we strongly recommend to see how to optimize a code with profilers, as $cProfiler$. Many codes can be used or even write from other languages and integrate with Python, not only to save calculation time, but also development time. See $Cython$ for $C/C++$, $Jython$ for Java, among others.

\bigskip
\bigskip

\textbf{6. CONCLUSIONS}

\bigskip
\bigskip

It is expected that this article has contributed to an overview of the Python programming language, and how it can be useful for scientists, particularly the astronomers. The relevance of such presentations is to equip community options for increasing use of computers with increasingly complex analysis and data volume.
 
Another point is that researchers need to plan since their daily programming activities, such as data reduction, even participate in decisions on large scientific projects. Often they do not have a solid background in programming, and high-level languages of great integrability can play an important role to reduce development time, not only the time spent by the researchers in the code, but also as the interface of different sectors of the project. I emphasize that this level of planning, such as with web applications, are beyond the scope of this work, but can be easy associated with the Python programming philosophy.

Along the text are shown features that make Python a unique language. So if the goal of a scientist is an easy to use, free and versatile programming language for the purpose of manipulation and visualization of data, with possible integration with other computing functionalities of interest, Python is undoubtedly one outstanding option.

\bigskip
\bigskip

{\bf ACKNOWLEDGMENTS: I thank Dr. Odair Gimenes, Guilherme Ferrari and Carlos Barbosa for useful discussions and tips.}


\begin{thebibliography}{43}
\bibitem{2007PLW} LUSZCZEK, P., \& DONGARRA, J.\ 2007, International Journal of High Performance Computing Applications, 21, 360 

\bibitem{2007ASPC..376..261M} MAGEE, D.~K., BOUWENS, R.~J., \& ILLINGWORTH, G.~D.\ 2007, Astronomical Data Analysis Software and Systems XVI, 376, 261 

\bibitem{2011JHEP...06..128A} ALWALL, J., HERQUET, M., MALTONI, F., MATTELAER, O., 
\& STELZER, T.\ 2011, Journal of High Energy Physics, 6, 128 

\bibitem{2007JOSAA..24.1580H} HOM, E.~F.~Y., MARCHIS, F., LEE, T.~K., HAASE, S., AGARD, D.~A., 
\& SEDAT, J.~W.\ 2007, Journal of the Optical Society of America A, 24, 1580 

\bibitem{2010A&A...524A..61N} NOORDAM, J.~E., \& SMIRNOV, O.~M.\ 2010, Astronomy and Astrophysics, 524, A61 

\bibitem{2008PASP..120..439C} COTTON, W.~D.\ 2008, Publications of the Astronomical Society of the Pacific, 120, 439 

\bibitem{sage} SAGE: Open Source Mathematics Software. 2011, $<$http://www.sagemath.org/$>$

\bibitem{pynbody} PYNBODY - A Python Toolkit for Astrophysical N-Body Simulation. 2011, $<$https://github.com/GuilhermeFerrari/PyNbody/$>$

\bibitem{enthought} ENTHOUGHT Python Distribution. 2011, $<$http://www.enthought.com/products/epd.php$>$

\bibitem{scisoft} ESO - The Scisoft Software Collection. 2011, $<$http://www.eso.org/sci/software/scisoft/$>$

\bibitem{STScI} STSCI - Space Telescope Science Institute: Software \& Hardware. 2011, $<$http://www.stsci.edu/resources/software\_hardware/$>$

\bibitem{scipy} SCIPY , 2011. $<$http://www.scipy.org/$>$

\bibitem{matplotlib} MATPLOTLIB: Python plotting, 2011. $<$http://matplotlib.sourceforge.net/$>$

\bibitem{tosi} TOSI, Sandro. Matplotlib for Python Developers, 2009. Packt Publishing.

\bibitem{PEP} PEP: Python Enhancement Proposals. 2011, $<$http://www.python.org/dev/peps/$>$

\bibitem{aplpy} APLPY Home Page. 2011, $<$http://aplpy.github.com/$>$

\bibitem{astrogrid} ASTROGRID, Virtual Observatory Software. 2011, $<$http://www.astrogrid.org/$>$

\end{thebibliography}
\end{document}